\documentclass[onecolumn,showpacs,prc,superscriptaddress,amsmath,amssymb,preprintnumbers]{revtex4-1}

\usepackage{amssymb}
\usepackage{amsfonts}
\usepackage{amsmath}
\usepackage{longtable}
\usepackage{epsfig}

\usepackage{xcolor}

\usepackage{lineno}

\usepackage[percent]{overpic}

\usepackage{hyperref}
\usepackage[all]{hypcap}
\hypersetup{
    colorlinks,
    citecolor=blue,
    filecolor=blue,
    linkcolor=black,
    urlcolor=blue
}

\def\pt{p_{\rm T}}
\def\av#1{\langle #1 \rangle}
\def 	\r2{\rho_2}

\def\vf{\varphi}

\begin{document}
\title{On feasibility of azimuthal flow studies with Principal Component Analysis}

\author{Igor Altsybeev}

\address{Saint-Petersburg State University, Universitetskaya nab. 7/9, St. Petersburg, 199034, Russia}

\email{Igor.Altsybeev@cern.ch}

\date{\today}

\begin{abstract}
It is shown 
that  the Principal Component Analysis applied to 
 azimuthal single-particle distributions
allows to perform  flow analysis in ways 
that are analogous to the traditional approaches based on multi-particle correlations.
In particular,
symmetric cumulants are considered.
%
It is demonstrated also  that 
statistical fluctuations due to a finite number of particles per event practically do not play a role for higher order  PCA-based cumulants.
\end{abstract}

\maketitle

\section{Introduction}

Principal Component Analysis (PCA)  is a method for decorrelation of multivariate data.
PCA  finds the most optimal basis for a given problem and thus reduces its dimensionality.
Recently,  it was suggested to apply PCA to heavy-ion collisions data 
on two-particle azimuthal correlations~\cite{Ollitrault_2015},
in order to reveal hidden patterns of the collective behaviour of the hadronic medium.
In \cite{IA_PCA} it was shown that 
 PCA  applied directly to single-particle azimuthal ($\vf$) distributions in A--A collisions 
reveals Fourier harmonics as the natural and the most optimal basis.

In the latter approach, 
technically, one needs to take 
distributions of particles in $M$ bins in each out of $N$ events, normalize
them and subtract the mean, and then apply PCA to the obtained $N$$\times$$M$ matrix (see more details 
in \cite{IA_PCA, Liu_2019}).
As an output from PCA, we have a set of  orthonormal eigenvectors (${\mathbf e}_i, i=1,...,M$), 
and also a set of coefficients $\alpha_i^k (k=1,...,N$) of PCA decomposition,
such that
the particle distribution in $k$-th event (denoted as  ${\mathbf x}^{(k)}$ that is a vector with $M$ elements)  can be written as
\begin{equation}
{\mathbf x}^{(k)} =  \sum_{i=1}^M   \alpha_i^{(k)}  {\mathbf e}_i\, .
\end{equation}
By construction, the first $K$ components ($K<M$)
contain the most of the total variance of a dataset.

The azimuthal flow  in heavy-ion collisions  is typically studied using expansion
of particle azimuthal probability density in a series: 
\begin{align}
\begin{split}
\label{flow}  
f(\vf)  = \frac{1}{2\pi}
\big[ 1 + 2 \sum_{n=1}^\infty v_n  \cos\big(n(\vf-\Psi_n)\big)  \big],
\end{split}
\end{align}
where $v_n$ are  the flow 
coefficients. 
As mentioned above,
PCA  applied to event-by-event  azimuthal  single-particle distributions
 reveals the Fourier basis, and thus coefficients of the PCA decomposition  gain a definite meaning.
If the decomposition \eqref{flow} is applied  event-by-event,
values of $\hat{v}_n$ observed in the $k$-th event
 are related to the PCA coefficients 
(assuming that the elliptic flow dominates, and the next is the triangular flow) 
as follows:
${\hat{v}_2 }^{(k)} =  \sqrt{M\over 2} \sqrt{ {\alpha_1^k}^2+{\alpha_2^k}^2}$,
${\hat{v}_3 }^{(k)} =  \sqrt{M\over 2} \sqrt{ {\alpha_3^k}^2+{\alpha_4^k}^2}$, and so on.

In this paper, it is demonstrated
 that the coefficients of the PCA decomposition can be combined
into  expressions that are
equivalent to the multi-particle cumulants
used in the flow studies.
The first observable considered in Section 2 
is the flow amplitude calculated via two- and four-particle correlations.
So-called symmetric cumulants
that measure correlations between amplitudes 
of flow harmonics of different orders
are studied in Section 3.
It is estimated also how the statistical fluctuations contribute 
to the values extracted using PCA.

\section{\bf Higher-order cumulants from PCA}

In flow studies,
the simplest way to get an estimate for the   amplitudes $v_n$
is to use the two-particle cumulants $c_n\{2\}$:
\begin{equation}
\label{vn_c2}
v_n\{2\} =   \sqrt{c_n\{2\}} =   \sqrt{  \av{ v_n^2  } } .
\end{equation}
It is well-known that this quantity suffers from the so called non-flow 
contributions coming from e.g.  resonance decays and jets.
Suppression of the non-flow 
is typically done 
by utilizing the multi-particle cumulants.
For example, 
the amplitude of the $n$-th harmonic can be estimated from 
the fourth-order cumulant $c_n\{4\}$
\cite{Ollitrault_et_al_2001} as
\begin{equation}
\label{vn_c4}
v_n\{4\} =   \sqrt[4]{-c_n\{4\}} =   \sqrt[4]{ 2 \av{ v_n^2}^2  - \av{ v_n^4} .
}
\end{equation}
We may try to adopt \eqref{vn_c2} and \eqref{vn_c4} 
for the flow studies with the PCA.
When one deals with event-by-event particle distributions (like in the present application of the PCA),
it is essential  to investigate the influence of the statistical fluctuations
due to a limited number of particles per event. 
Following the approach used, for instance,  in \cite{Jia:2014vja, He_Qian_Huo:2017},
we denote a {\it true} amplitude of the $n$-th Fourier harmonic in a given event as $v_n$
and an amplitude of the statistical noise as $a_n$.
If $\hat{v}_n$ is the {\it observed} amplitude,  extracted by PCA in a given event,
%
projections of the  corresponding flow 
vector 
on  $x$ and $y$ axes in the transverse plane are
\begin{equation}
\hat{v}_{n,x} =  v_{n,x}+a_{n,x} ,   \hspace{0.5cm}
\hat{v}_{n,y} =  v_{n,y}+a_{n,y}  .
\end{equation}
The squares 
of the $v_n$, $a_n$ and $\hat{v}_n$ are 
\begin{equation}
	v_n^2 =  v_{n,x}^2 + v_{n,y}^2 ,
\end{equation}
\begin{equation}
	a_n^2 =  a_{n,x}^2 + a_{n,y}^2 ,
\end{equation}
\begin{equation}
\label{vn_hat_squared}
\hat{v}_n^2 =  \hat{v}_{n,x}^2 + \hat{v}_{n,y}^2
= v_n^2  + a_n^2  + 2 ( v_{n,x} a_{n,x} +  v_{n,y} a_{n,y} ) .
\end{equation}
After averaging \eqref{vn_hat_squared}
over events, we get
\begin{equation}
\label{av_vn_hat_squared}
	\av{\hat{v}_n^2}
		= \av{v_n^2}  + \av{a_n^2 } 
		+ 2 \big< v_{n,x} a_{n,x} \big> + 2  \big<v_{n,y} a_{n,y} \big>.
\end{equation}
If we assume that signal and the statistical noise are uncorrelated,
the last two terms factorize:
$\big< v_{n,x} a_{n,x} \big> = \av{ v_{n,x}} \av{ a_{n,x} }$
and
$\big< v_{n,y} a_{n,y} \big> = \av{ v_{n,y}} \av{ a_{n,y} }$.
Since event-averaged values of the $x$- and $y$-components
of $v_n$ and $a_n$ are zero,
\eqref{av_vn_hat_squared}
becomes
\begin{equation}
\label{av_vn_hat_squared_without_last_terms}
	\av{\hat{v}_n^2}
		= \av{v_n^2}  + \av{a_n^2 } ,
\end{equation}
and the true value $v_n$ is  found by inverting \eqref{av_vn_hat_squared_without_last_terms}:
\begin{equation}
\label{av_vn_truth}
	\av{v_n^2}  = \av{\hat{v}_n^2} -  \av{a_n^2 } .
\end{equation}
This result was obtained in  \cite{IA_PCA}
and  can be used to get an estimation of the Fourier amplitudes
 $v_n\{2\}$ using \eqref{vn_c2}. 
Values $\av{a_n^2 }$ measure the statistical fluctuations. They
can be calculated
 by applying PCA to the same events, but with  randomized $\vf$-angles.

\begin{figure}[t]
\centering
\begin{overpic}[width=0.6\textwidth, trim={0.1cm 0.0cm 1.3cm 1.4cm},clip]
{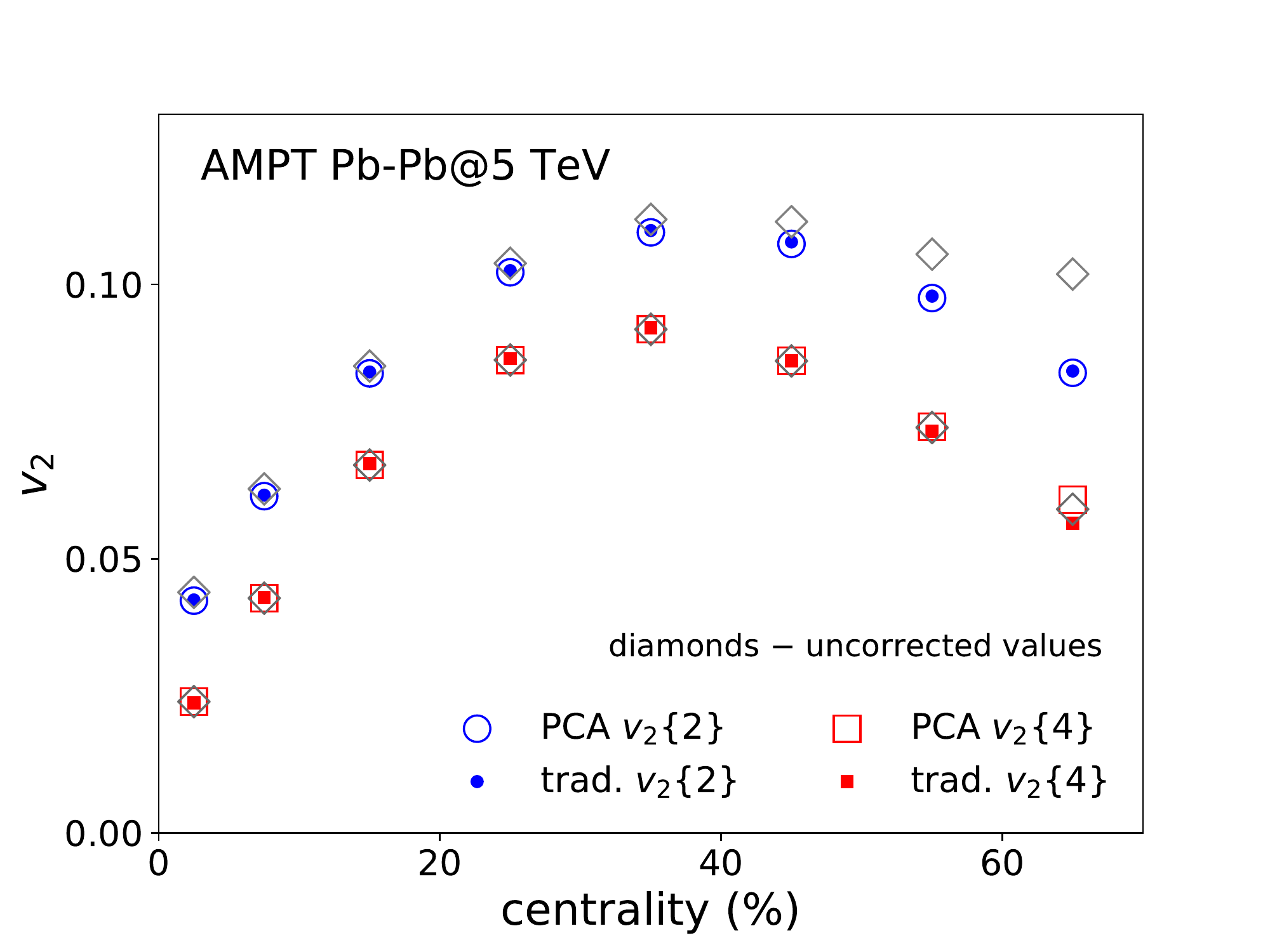} 
\end{overpic}  
\vspace{-2mm}
\caption{
Amplitudes $v_2$ of the  second Fourier  harmonic 
as a function of centrality 
in Pb-Pb collisions at 5~TeV in AMPT (2.8$\times 10^6$ events). 
Values  extracted by PCA are shown by open markers: circles --  $v_2\{2\}$,
squares --  $v_2\{4\}$.
Small full markers show calculations 
using standard 2- and 4-particle cumulant methods.
Gray diamonds denote PCA values before the correction on the statistical noise.
Analysis is performed for charged particles with pseudorapidity  $|\eta|<0.8$
within transverse momentum ($\pt$) range is 0.2--5 GeV/$c$.
Number of $\varphi$ bins used for PCA is $M=48$.
}
\label{AMPT_2pc_vs_4pc}
\end{figure}

The effect from the statistical noise correction on the $v_2$  
is shown   in Figure~\ref{AMPT_2pc_vs_4pc} 
for  Pb-Pb events simulated in
AMPT  generator (2.8$\times 10^6$ events).
Uncorrected raw $\hat{v}_2$ values  (upper gray diamonds), extracted directly from PCA,
after the correction 
become blue open circles.
These circles are on top of the values obtained with the traditional two-particle cumulant method (full  circles). It can be seen that effect from the correction
is more pronounced for the peripheral events,
where a number of particles per event is lower.

For the fourth power of the observed magnitude, 
we can use \eqref{vn_hat_squared} again:
\begin{equation}
\label{vn4_hat}
\hat{v}_n^4 =  \big[
v_n^2  + a_n^2  + 2 ( v_{n,x} a_{n,x} +  v_{n,y} a_{n,y} ) 
\big]^2 .
\end{equation}
Averaging \eqref{vn4_hat} over events
and taking into account that 
$x$ and $y$ components of the noise are independent and their variances are equal,
$\av{a_{n,x} a_{n,y}} = \av{a_{n,x} } \av{a_{n,y}  }$
and $\av{ a_{n,x}^2 } = \av{ a_{n,y}^2 }$,
we get 
\begin{equation}
\label{vn4_hat_averaged}
\av{ \hat{v}_n^4 } =   
\av{v_n^4}  + \av{a_n^4}  + 4 \av{v_n^2} \av{a_n^2}.
\end{equation}
From \eqref{av_vn_truth} and \eqref{vn4_hat_averaged}, 
\begin{equation}
\label{vObs4}
2\av{ \hat{v}_n^2 }^2 - \av{ \hat{v}_n^4 }  =   
2\av{v_n^2}^2 - \av{v_n^4}  + 2 \av{a_n^2}^2
- \av{a_n^4}  ,
\end{equation}
and, inverting \eqref{vObs4}
and
 substituting into \eqref{vn_c4},
we get the following 
 estimation for the $v_n$: 
\begin{equation}
\label{vn_cum4_corrected}
v_n\{4\}=   
\sqrt[4]{
  2\av{ \hat{v}_n^2}^2  - \av{ \hat{v}_n^4}
- \big( 2 \av{a_n^2}^2  - \av{a_n^4}  \big) .
}
\end{equation}
From this expression,
one may note that when $v_n \gg a_n$,
values of $v_n\{4\}$ are remarkably insensitive to  the statistical noise.
Indeed, for a 
hypothetical case of constant magnitudes of the flow and the statistical noise, 
 $v_n\{4\} = \sqrt[4]{ \hat{v}_n^4 - a_n^4}$,
while $v_n\{2\} = \sqrt{ \hat{v}_n^2 - a_n^2}$.

In Figure \ref{AMPT_2pc_vs_4pc},
open squares 
correspond to estimations of $v_n\{4\}$ based on PCA coefficients,
formula \eqref{vn_cum4_corrected}.
They match small closed squares that stand for $v_n\{4\}$
calculated with the traditional approach using four-particle correlations.
At the same time, it can be seen
that the raw values $\hat{v}_n\{4\}$ (diamonds), calculated by \eqref{vn_cum4_corrected}
without taking into account the statistical term,
 are almost on top of the squares even for peripheral events.
This indicates that 
the correction for statistical fluctuations is almost irrelevant for the  $v_n\{4\}$
as soon as the number of particles in events is large enough.
Toy studies showed that this is the case when the flow  magnitude is $v_n\sim 0.1$
and a number of particles is $\gtrsim$100.

\section{Symmetric cumulants with PCA}

Following the same strategy,
we can investigate 
feasibility of the  studies 
of the  so-called symmetric cumulants \cite{Bilandzic_2014}
with the PCA. 
This observable measures
correlations between the amplitudes of the $n$-th and $m$-th
Fourier harmonics and is defined as 
\begin{equation}
\label{SC_nm}
{\rm SC}(n,m)
= \av{v_n^2  v_m^2}  
-  \av{v_n^2} \av{ v_m^2}.
\end{equation}
Previous attempt to study 
the symmetric cumulants with the PCA
was done in a paper \cite{Liu_2019}.
Since the PCA basis  obtained in \cite{Liu_2019} was somehow distorted
(i.e.  not identical to the Fourier harmonics),
extracted SC in this paper values did not match
the ``truth'' values. 

We start with the single-event raw quantity:
\begin{multline}
\label{vnvm_hat}
\hat{v}_n^2  \hat{v}_m^2 
=  
( \hat{v}_{n,x}^2 + \hat{v}_{n,y}^2)
(\hat{v}_{n,x}^2 + \hat{v}_{n,y}^2
) = \\
=
\big[ ( v_{n,x} + a_{n,x} )^2 + ( v_{n,y} + a_{n,y} )^2 \big]
\big[ ( v_{m,x} + a_{m,x} )^2 + ( v_{m,y} + a_{m,y} )^2 \big] = \\
= \big[ v_n^2  + a_n^2  + 2 ( v_{n,x} a_{n,x} +  v_{n,y} a_{n,y} ) \big]
\big[ v_m^2  + a_m^2  + 2 ( v_{m,x} a_{m,x} +  v_{m,y} a_{m,y} ) \big].
\end{multline}
Averaging over events and taking into account
$\av{v_{n,x}}=\av{v_{n,y}}=\av{a_{n,x}}=\av{a_{n,y}}=0$ (the same for the $m$-th harmonic), and also the factorization of the noise harmonics $\av{a^2_n a^2_m} = \av{a^2_n}\av{ a^2_m}$
as well as the noise and the signal $\av{v^2_n a^2_m} = \av{v^2_n}\av{ a^2_m}$, 
we obtain
\begin{equation}
\label{vnvm_hat_averaged}
\av{\hat{v}_n^2  \hat{v}_m^2} 
=  \av{v_n^2  v_m^2}  
+ \av{v_n^2}\av{ a_m^2}   + \av{v_m^2}\av{ a_n^2} 
+ \av{a_n^2}\av{ a_m^2} .
\end{equation}
Using \eqref{av_vn_truth} and  \eqref{vnvm_hat_averaged}, 
the desired term $\av{v_n^2  v_m^2}$ 
can be expressed via ``measurable'' quantities as
\begin{equation}
\label{vnvm_averaged}
 \av{v_n^2  v_m^2}
=  \av{\hat{v}_n^2  \hat{v}_m^2} 
- \av{\hat{v}_n^2}\av{ a_m^2}     
- \av{\hat{v}_m^2}\av{ a_n^2}   
 +\av{a_n^2}\av{ a_m^2} .
\end{equation}
The final expression for the ${\rm SC}(n,m)$ is thus
\begin{equation}
\label{SC_nm_corrected}
{\rm SC}(n,m)
= \av{v_n^2  v_m^2}  
-  \av{v_n^2} \av{ v_m^2} 
=  \av{\hat{v}_n^2  \hat{v}_m^2}  
-  \av{\hat{v}_n^2} \av{ \hat{v}_m^2} .
\end{equation}
It is remarkable that all terms related to the statistical noise are canceled.

\begin{figure}[t]
\centering
\begin{overpic}[width=0.6\textwidth, trim={0.1cm 0.0cm 1.3cm 1.cm},clip]
{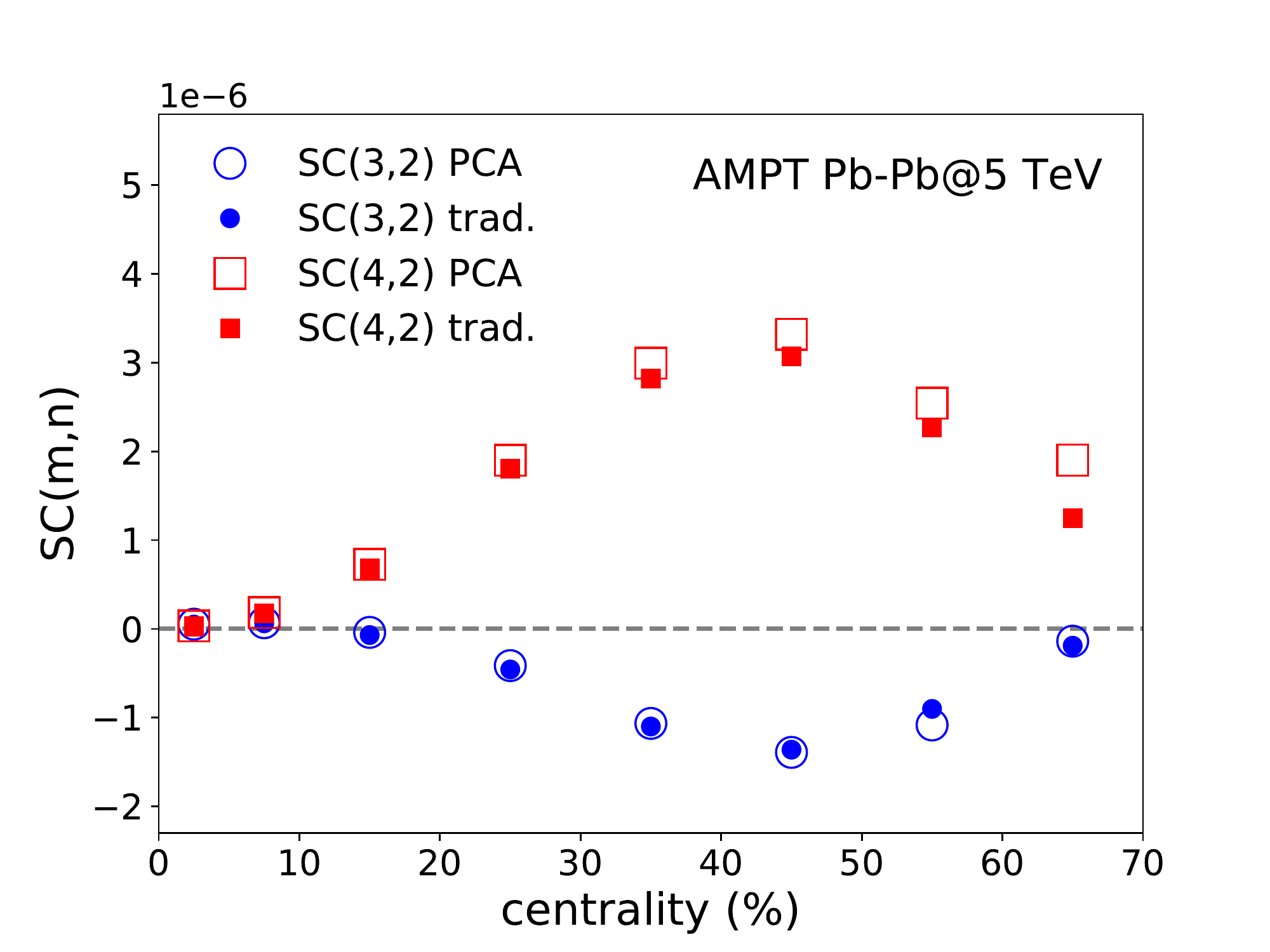} 
\end{overpic}  
\vspace{-2mm}
\caption{
Centrality dependence 
of the symmetric cumulants SC(3,2) and SC(4,2) in Pb-Pb collisions from AMPT.
Open markers -- values are extracted from PCA,
closed markers -- by traditional method of multi-particle correlations. 
Particles within  $|\eta|<2.4$, $\pt$ range 0.2--5 GeV/$c$.
}
\label{AMPT_SC}
\end{figure}

Figure \ref{AMPT_SC} shows
centrality dependence 
of the symmetric cumulants SC(3,2) and SC(4,2) in Pb-Pb collisions from AMPT.
It can be seen that  values  extracted from PCA (open markers)
match with calculations using multi-particle correlations (closed markers).
For SC(4,2), there are slight deviations in peripheral centrality classes,
a possible reason is 
the event-plane 
correlation of these two harmonics. Detailed investigation of this is out of the scope of this  article.

\section{ Summary }

It was shown that 
 the Principal Component Analysis applied to event-by-event
azimuthal single-particle distributions
allows to perform  flow analyses 
that are analogous to the traditional approaches based on multi-particle correlations. 
As the first example,  flow amplitudes based on the fourth-order cumulant were considered. 
As the second case,
correlations between flow amplitudes 
in terms of symmetric cumulants were calculated.
Using realistic events from the  AMPT generator,
PCA results were directly compared to calculations using the traditional 
techniques,  a good correspondence was obtained.
It was demonstrated also that a contribution 
from statistical fluctuations due to a finite number of particles per event 
to the higher-order PCA-based cumulants is small.
%

\section*{\bf Acknowledgements }
This study is supported  by Russian Science Foundation, grant 17-72-20045.


\bibliography{bibliography}

\end{document}